\documentclass[conference]{IEEEtran}
\IEEEoverridecommandlockouts

\usepackage{cite}
\usepackage{amsmath,amssymb,amsfonts}
\usepackage{algorithm}
\usepackage{algpseudocode}
\usepackage{graphicx}
\usepackage{booktabs}
\usepackage{multirow}
\usepackage{xcolor}
\usepackage[table]{xcolor}
\usepackage{siunitx}
\usepackage[caption=false,font=footnotesize]{subfig}
\usepackage{hyperref}
\usepackage{placeins}
\usepackage{float}
\usepackage{tabularx}
\usepackage{colortbl}
\usepackage{tikz}

\def\BibTeX{{\rm B\kern-.05em{\sc i\kern-.025em b}\kern-.08em
    T\kern-.1667em\lower.7ex\hbox{E}\kern-.125emX}}
\begin{document}

\title{Embedding 
Single-Phase Grid-Forming Inverters 
in Three-Phase  Unbalanced 
Power Flow

\thanks{This work relates to the Department of Navy award N000142512374 issued by the Office of Naval Research. \textit{Corresponding Author: Peng Zhang.}}
}

\author{
    \IEEEauthorblockN{Kamini Shahare, Fei Feng, 
    and Peng Zhang}
    \thanks{
         K. Shahare and P. Zhang are with the Department of Electrical and Computer Engineering, Stony Brook University, Stony Brook, NY, 11794-2350, USA (e-mail: kamini.shahare, P.Zhang@stonybrook.edu). } 
        \thanks{    F. Feng is with the State University of New York Maritime College, Bronx, NY 10465, USA (e-mail: ffeng@sunymaritime.edu).
    }
    \
}
\maketitle

\begin{abstract}
This letter introduces Grid-Forming Three-Phase Power Flow (GFM-3PF), an augmented Newton power flow framework for modeling single-phase grid-forming inverters (GFMs) in three-phase unbalanced networks. The contributions of this work are threefold: 1) developing a two-stage solution strategy based on positive-sequence initialization followed by three-phase phase-domain lifting for robust large-scale computation; 2) incorporating phase-domain unbalanced network modeling with controlled mutual coupling together with phase-dependent, frequency-sensitive load representation; and 3) embedding a single-phase GFM bus model directly into the three-phase unbalanced Newton power flow equations through a common 
droop-governed frequency state. GFM-3PF is validated on 3-bus, 33-bus, and 118-bus test systems, demonstrating its accuracy, robustness, and scalability.

\end{abstract}

\section{Introduction}
Conventional power flow formulations are increasingly challenged by inverter-based resources, three-phase unbalance, and reduced synchronous generation \cite{b1}. Grid-forming inverters (GFMs) 
in low-inertia systems are crucial because their steady-state operating point is governed by droop control and cannot be correctly 
represented by standard slack or PV buses 
\cite{b2}. Although existing three-phase 
power flow methods capture asymmetric loading and phase coupling \cite{b3}, a unified formulation that embeds droop-controlled single-phase GFM behavior into three-phase network equations remains needed \cite{b4}.

This letter presents Grid-Forming Three-Phase Power Flow (GFM-3PF), an augmented Newton framework that 
enforces the single-phase GFMs 
droop equations directly in the Newton residual. Unlike conventional three-phase power flow with fixed frequency, GFM-3PF solves the phase-domain voltage profile and droop-governed frequency simultaneously, capturing the coupling among unbalanced loading, single-phase GFM injection, and frequency-sensitive demand.

\section{GFM-3PF Modeling}
The proposed GFM-3PF formulation solves the three-phase unbalanced steady-state operating point of a network 
interconnected with single-phase GFMs. 
Unlike a conventional slack-bus formulation, the common system frequency is treated as an additional unknown and determined by the active-power/frequency droop relation of the participating single-phase GFMs. One GFM phase angle is selected only as the mathematical angular reference, while the phase voltages and common frequency are solved simultaneously within an augmented Newton framework.

\subsection{Single-Phase GFM Bus Formulation}

Consider a set of single-phase 
GFMs denoted by
$\mathcal{G}=\{(b_i,\phi_i,\gamma_i,P_i^\star,V_i^\star)\}_{i=1}^{n_g}$,
where $b_i$ is the bus index, $\phi_i\in\{A,B,C\}$ is the connected phase, $\gamma_i$ is the participation factor, $P_i^\star$ is the scheduled active-power setpoint, and $V_i^\star$ is the voltage-magnitude setpoint of the $i$th GFM. Unlike a conventional slack bus,
the GFM does not prescribe both system frequency and active-power injection.
Instead, its steady-state operating point is determined by the network
equations together with an active-power/frequency droop relation.

For a single-phase GFM connected to phase $\phi_i$ of bus $b_i$, the droop
power is defined using 
\begin{equation}
P_{\mathrm{droop},i}=P_{b_i,\phi_i}.
\end{equation}
To avoid tying frequency deviation to absolute injected power, a scheduled
or nominal active-power setpoint $P_i^\star$ is included. The aggregate
droop power is therefore defined as
\begin{equation}
P_{\mathrm{agg}} =
\sum_{i=1}^{n_g} \gamma_i \left(P_{b_i,\phi_i}-P_i^\star\right).
\end{equation}
The common per-unit frequency is then governed by
\begin{equation}
\omega = \omega_0 - m_p P_{\mathrm{agg}},
\end{equation}
where $\omega_0=1$ pu is the nominal per-unit frequency and $m_p$ is the
per-unit active-power/frequency droop coefficient.

In the augmented Newton formulation, this droop relation is enforced through
the frequency residual
\begin{equation}
\Delta W =
\omega_0
- m_p \sum_{i=1}^{n_g}\gamma_i\left(P_{b_i,\phi_i}-P_i^\star\right)
-\omega .
\end{equation}
For a single GFM, this expression reduces directly to the droop relation of
the connected phase. For multiple single-phase GFMs, the same equation
captures their aggregate contribution through the participation factors
$\gamma_i$.

In addition to active-power/frequency droop, the GFM-connected phase is
modeled as a voltage-regulating phase with a specified voltage-magnitude
setpoint,
\begin{equation}
|V_{b_i,\phi_i}| = V_i^\star .
\end{equation}
Thus, the GFM phase regulates its voltage magnitude, while its phase angle is
solved as part of the network equations, except for one selected GFM phase
angle used as the mathematical reference. The reactive-power injection
$Q_{b_i,\phi_i}$ is not specified a priori; it is obtained from the
phase-domain network solution required to satisfy the voltage-magnitude
constraint and power balance. In this letter, the focus is on the steady-state
impact of active-power/frequency droop; explicit reactive-power/voltage droop
and inverter current or power limits are not included and are left for future
extensions.

\subsection{Three-Phase Unbalanced Network Modeling}
A positive-sequence power flow is first solved to provide an initial operating point. The positive-sequence voltage $V^{(+)}$ is then expanded to phase quantities as
\begin{equation}
V_a = V^{(+)}, \qquad V_b = a^2 V^{(+)}, \qquad V_c = aV^{(+)}
\end{equation}
where $a=e^{j2\pi/3}$ is the standard three-phase rotation operator. In this letter, the phase-domain admittance matrix is constructed using self terms and a controlled mutual-coupling factor $\mu$ for the synthetic test systems. This provides a consistent way to evaluate the proposed augmented Newton formulation under unbalanced and coupled phase conditions. For practical feeder studies, the same formulation can directly use measured or benchmark phase-domain line impedance matrices.

\subsection{Unbalanced Load Modeling}
To represent phase unbalance, the total demand at each bus is distributed
using phase participation coefficients
$\boldsymbol{\alpha}=[\alpha_A,\alpha_B,\alpha_C]$.
The active and reactive loads are modeled using general 
scaling functions $s_P(\omega)$ and $s_Q(\omega)$ as
\begin{equation}
P_{D,\phi}=s_P(\omega)\alpha_\phi P_D, \qquad
Q_{D,\phi}=s_Q(\omega)\alpha_\phi Q_D ,
\end{equation}
where $\phi\in\{A,B,C\}$. This formulation is not restricted to a particular
load-frequency model; linear, exponential, or other nonlinear load-frequency
characteristics can be incorporated by selecting the corresponding
$s_P(\omega)$ and $s_Q(\omega)$ and including their derivatives in the
augmented Jacobian. In the case studies, a linear approximation around
$\omega_0=1$ pu is used as an illustrative example: \vspace{-4pt}
\begin{equation}\vspace{1pt}
s_P(\omega)=1+k_{pf}(\omega-\omega_0), \qquad
s_Q(\omega)=1+k_{qf}(\omega-\omega_0). \vspace{4pt}
\end{equation} 

\section{Augmented Newton Solution}

The proposed method solves the phase-domain network equations together with
the GFM droop equation in a single augmented Newton system. To clarify the
closure of the formulation, Table~\ref{tab:bus_type} summarizes the known
quantities, unknown variables, and residual equations used for each
bus/phase type.

\begin{table}[!t]
\centering
\caption{Bus/phase modeling structure in the proposed formulation} \vspace{-4pt}
\label{tab:bus_type}
\scriptsize
\renewcommand{\arraystretch}{1.15}
\setlength{\tabcolsep}{2.5pt}
\begin{tabular}{p{0.19\linewidth} p{0.23\linewidth} p{0.24\linewidth} p{0.24\linewidth}}
\hline
\hline
\textbf{Bus/phase type} & \textbf{Known quantities} & \textbf{Unknowns} & \textbf{Residual equations} \\
\hline
PQ phase &
$P_D$, $Q_D$ &
$V_r$, $V_i$ &
$\Delta P$, $\Delta Q$ \\
\hline
Reference phase &
$\angle V_{\mathrm{ref}}$ &
Compatible source injection &
Reference-angle constraint \\
\hline
Single-phase GFM phase &
$V_i^\star$, $m_p$, $P_i^\star$, $\gamma_i$ &
Phase voltage state, $Q_{b_i,\phi_i}$, common $\omega$ &
$\Delta P$, voltage-magnitude constraint, $\Delta W$ \\
\hline
Non-GFM phase at GFM bus &
Assigned load or network injection &
$V_r$, $V_i$ &
$\Delta P$, $\Delta Q$ \\
\hline
\hline
\end{tabular}
\end{table}

Let $\mathcal{N}_{nr}$ denote the set of non-reference phase nodes whose
voltages are solved in the Newton iteration. The unknown vector is defined as
\begin{equation}
\mathbf{x} =
\begin{bmatrix}
\Re(\mathbf{V}_{\mathcal{N}_{nr}}) \\
\Im(\mathbf{V}_{\mathcal{N}_{nr}}) \\
\omega
\end{bmatrix}.
\end{equation}
Here, $\mathbf{V}_{\mathcal{N}_{nr}}$ contains the complex phase-domain
voltages of all solved non-reference phase nodes, including PQ phases,
non-GFM phases at GFM buses, and GFM-connected phases whose voltage magnitude
is constrained by $V_i^\star$. The angle of one selected GFM phase is removed
from the unknown vector and used only as the mathematical reference.

The residual vector combines the phase-domain active-power mismatch, the
reactive-power or voltage-magnitude constraints, and the GFM droop-frequency
residual:
\begin{equation}
\mathbf{r} =
\begin{bmatrix}
\Delta \mathbf{P} \\
\Delta \mathbf{g} \\
\Delta W
\end{bmatrix}.
\end{equation}
Here, $\Delta \mathbf{g}$ denotes $\Delta Q$ for PQ phases and the
voltage-magnitude residual $|V_{b_i,\phi_i}|-V_i^\star$ for GFM-connected
phases. The corresponding reactive-power injection of the GFM phase is
obtained from the network solution. The scalar residual $\Delta W$ enforces
the aggregate single-phase GFM droop relation. After selecting one angular
reference, the number of residual equations equals the number of unknowns in
the augmented Newton system.

The nonlinear equations are solved iteratively using
\begin{subequations}
\begin{align}
\mathbf{J}_{\mathrm{aug}}\Delta \mathbf{x} &= -\mathbf{r}(\mathbf{x}), \\
\mathbf{x}^{k+1} &= \mathbf{x}^{k} + \Delta \mathbf{x},
\end{align}
\end{subequations}
where $\mathbf{J}_{\mathrm{aug}}=\partial\mathbf{r}/\partial\mathbf{x}$ is the
augmented Jacobian. Compared with a conventional three-phase power-flow
Jacobian, $\mathbf{J}_{\mathrm{aug}}$ includes additional derivatives
associated with frequency-sensitive load scaling, the GFM voltage-magnitude
constraint, and the droop-frequency residual. Therefore, the phase voltages
and common-frequency state are updated simultaneously within one Newton
iteration.

\section{Test Case Scenarios}
To demonstrate the performance and scalability of GFM-3PF, three systems of different scales are thoroughly tested.  
\subsection{IEEE 3-bus Test Feeder}
\begin{figure}[t]
    \centering
    \begin{tikzpicture}
        \node[inner sep=0pt] {\includegraphics[width=0.48\textwidth]{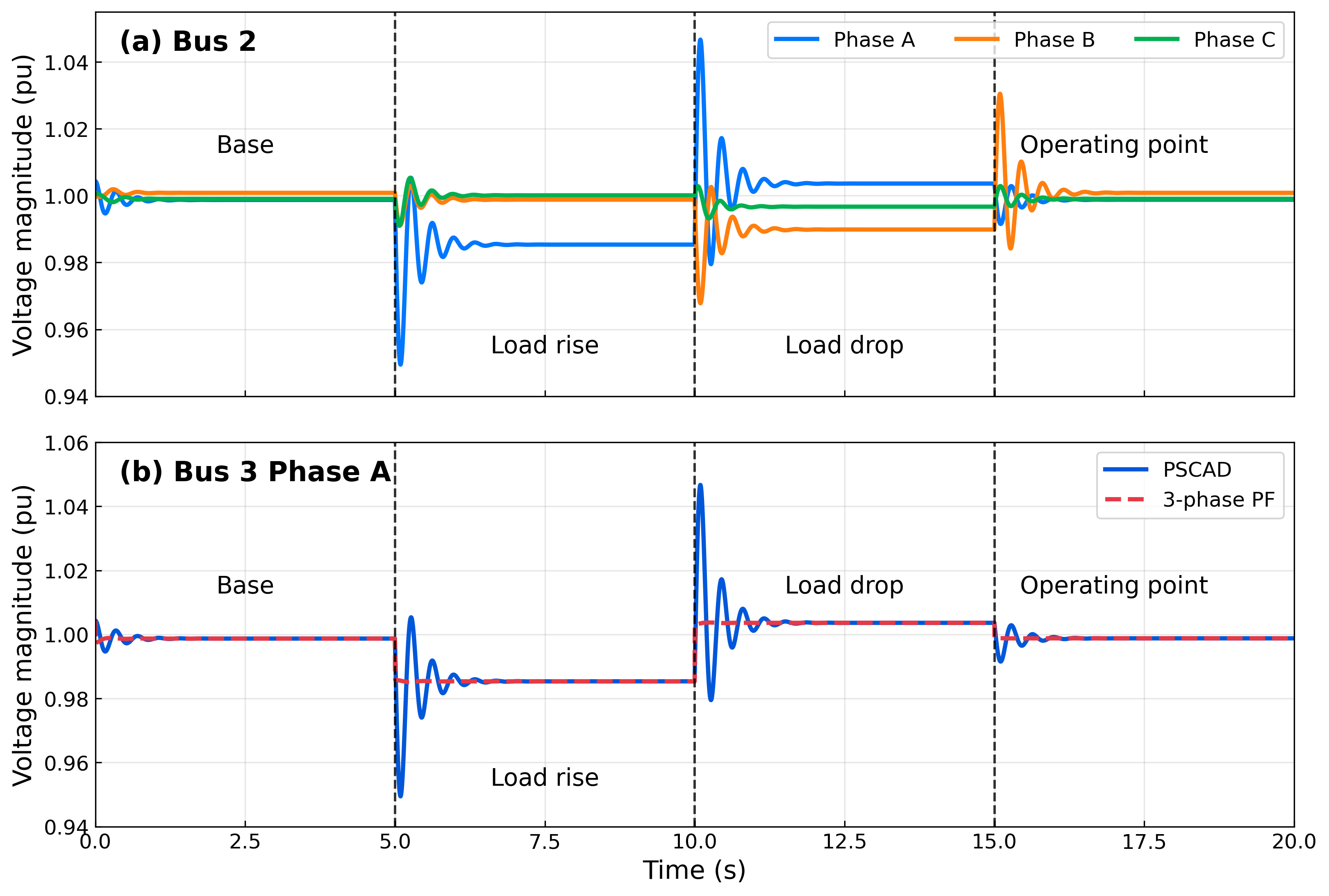}};
    \end{tikzpicture}
    \vspace{-0.3cm}
    \caption{Three-phase voltage magnitude profile under different disturbance events at (a) Bus~2 all phases (b) Bus 3 phase A comparison between GFM-3PF and PSCAD .}
    \label{fig1}
\end{figure}

A 3-bus system with a source at Bus~1, a PQ load bus at Bus~2, and a single-phase GFM at Bus~3 is used to verify the proposed framework. Fig.~\ref{fig1}(a) shows the bus voltage magnitudes, with VUF below 3\% at all buses, while Fig.~\ref{fig1}(b) shows the comparison of GFM-3PF with PSCAD under events at 5~s, 10~s, and 15~s. PSCAD captures sharper EMT transients, whereas GFM-3PF reproduces the associated operating-point transitions and closely matches the pre-event and post-event voltage levels.

\subsection{IEEE 33-bus System}
The IEEE 33-bus case is used to assess the proposed method on a medium-scale radial feeder. In this case, the single-phase GFM is placed at bus~4 on phase A, consistent with the phase-selective GFM configuration summarized in Table~\ref{tab:comp_summary}. The solved common frequency is
$\omega=0.999807$ pu after applying the phase-specific droop model with the active-power setpoint $P_i^\star$. The final mismatch is $4.24\times10^{-8}$, confirming convergence of the proposed augmented Newton formulation.

\begin{figure}[t]
    \centering
    \begin{tikzpicture}
        \node[inner sep=0pt] {\includegraphics[width=0.48\textwidth]{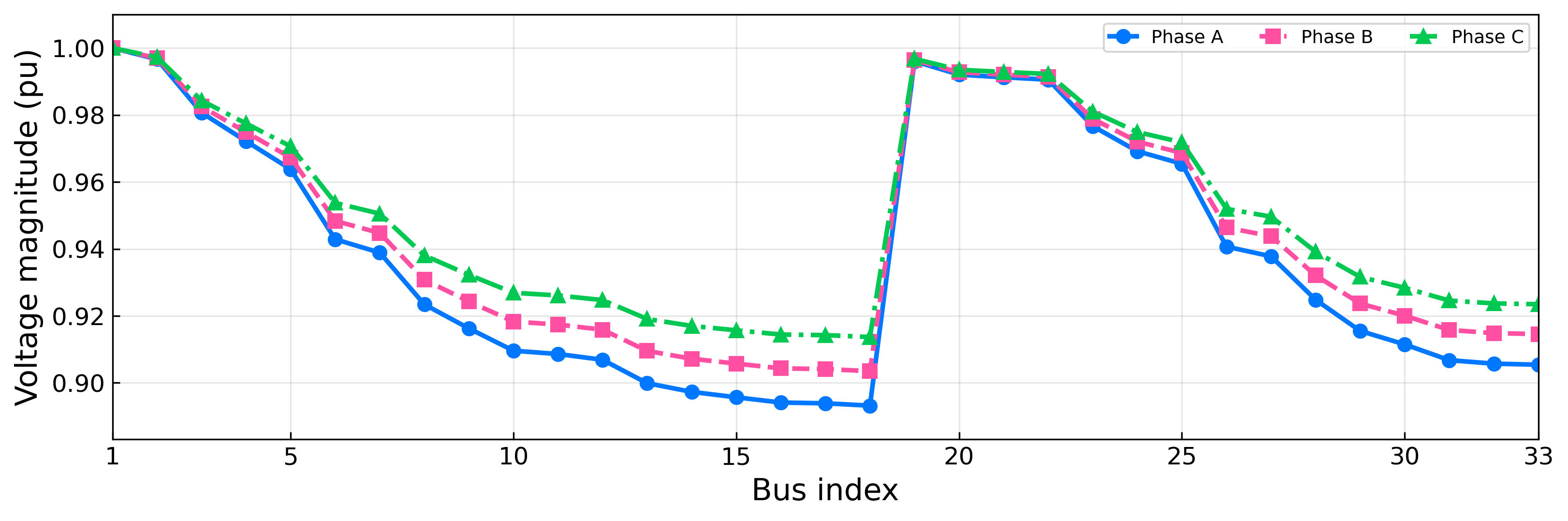}};
    \end{tikzpicture}
    \caption{Three-phase Voltage magnitude profile for IEEE 33-bus radial distribution System.}
    \label{fig3}
\end{figure}

Fig.~\ref{fig3} shows the three-phase voltage magnitude profile of the unbalanced 33-bus feeder. The phase-dependent separation among the voltage trajectories reflects the imposed unbalanced loading, while the nodal voltage attenuation along the feeder is consistent with radial distribution-system behavior.

\subsection{IEEE 118-bus Meshed System}
\begin{figure}[t]
    \centering
    \begin{tikzpicture}
        \node[inner sep=0pt] {\includegraphics[width=0.48\textwidth]{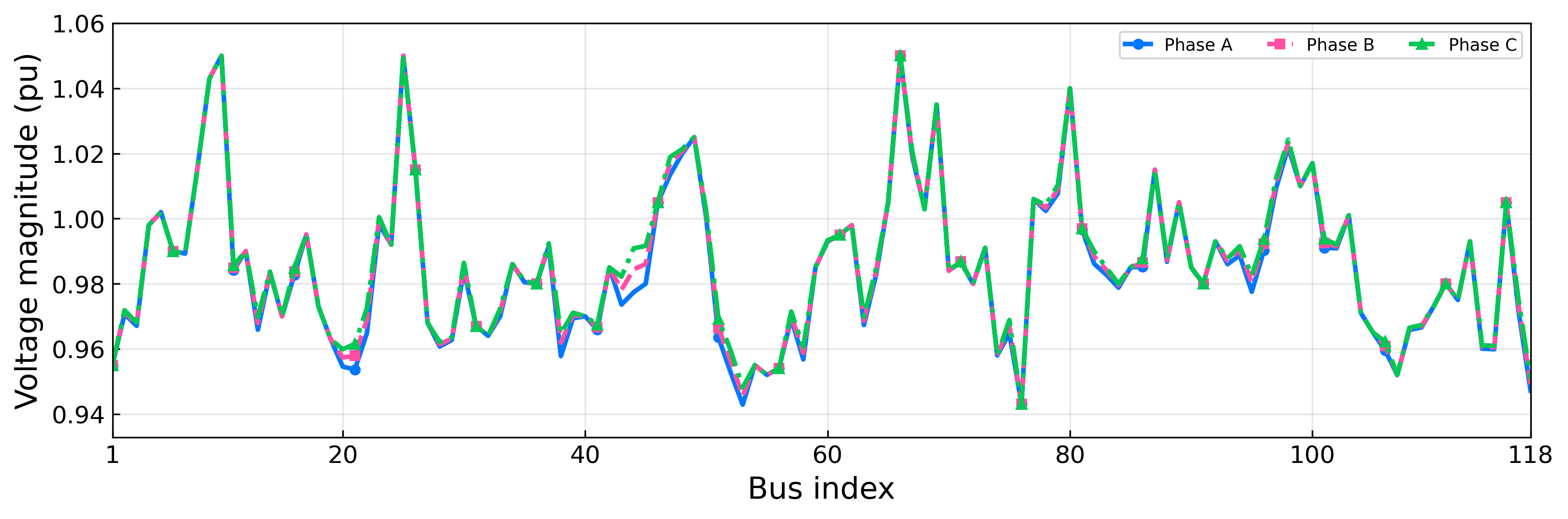}};
    \end{tikzpicture}
    \caption{Three-phase voltage magnitude profile for IEEE 118-bus meshed System.}
    \label{fig4}
\end{figure}

The IEEE 118-bus case is used to demonstrate the scalability of GFM-3PF in a large-scale meshed network. Two single-phase GFMs are placed on different phases to provide phase-selective support: one at Bus~69 on phase A and another at Bus~1 on phase B. The corresponding single-phase GFM ratings are 1.00 pu and 1.02 pu, respectively, on the adopted system base. The three-phase network includes phase mutual coupling, and the loads are distributed unevenly across phases with frequency-sensitive scaling. The solved common frequency is $\omega=0.99861$ pu, and the final mismatch is $4.854\times10^{-7}$, satisfying the prescribed tolerance of $10^{-6}$. These results confirm that GFM-3PF maintains stable convergence for the large-scale meshed case with phase-selective single-phase GFMs. Fig.~\ref{fig4} shows the three-phase voltage magnitude profile of the IEEE 118-bus meshed system.

\begin{table}[!t]
    \centering
    \caption{Performance Summary of GFM-3PF.}
    \label{tab:comp_summary}
    \scriptsize
    \setlength{\tabcolsep}{3pt}
    \begin{tabular}{c|c|c|c}
        \hline
        \hline
        \textbf{Item} & \textbf{3-Bus} & \textbf{33-Bus} & \textbf{118-Bus} \\ \hline
        GFM location & Bus 3-A & Bus 4-A & Bus 1-B, Bus 69-A \\ \hline
        $\omega$ (pu) & 0.99902 & 0.999807 & 0.99861 \\ \hline
        Iterations & 4 & 19 & 20 \\ \hline
        Final mismatch & $9.14\times10^{-13}$ & $4.240\times10^{-8}$ & $4.854\times10^{-7}$ \\ \hline
        Tolerance & $10^{-6}$ & $10^{-6}$ & $10^{-6}$ \\ \hline
        Key point & Verification & Medium-scale radial & Multi-GFM scalable \\ \hline
        \hline
    \end{tabular}
\end{table}

\section{Conclusion} 
This letter presented GFM-3PF, a Newton-based framework for embedding single-phase GFM behavior within a three-phase unbalanced power flow solver. GFM-3PF integrates positive-sequence initialization, three-phase unbalanced network modeling, frequency-sensitive load representation, and a
droop-governed common per-unit frequency state within a unified steady-state framework. Results on 3-bus, IEEE 33-bus, and IEEE 118-bus systems demonstrate the accuracy, robustness, and scalability of the method, showing its potential as a practical tool for steady-state analysis of inverter-dominated unbalanced power systems. \\


\begin{thebibliography}{00}
\bibitem{b1} P.~A.~N. Garcia, J.~L.~R. Pereira, S. Carneiro, Jr., V.~M. da Costa, and N. Martins,
``Three-Phase Power Flow Calculations Using the Current Injection Method,''
\textit{IEEE Transactions on Power Systems}, vol.~15, no.~2, pp.~508--514, May 2000.

\bibitem{b2} F. Feng and P. Zhang, "Enhanced Microgrid Power Flow Incorporating Hierarchical Control," \textit{IEEE Transactions on Power Systems}, vol. 35, no. 3, pp. 2463-2466, May 2020.


\bibitem{b3} D. Li, Y. Su, F. Wang, M. Olama, B. Ollis and M. Ferrari, "Power Flow Models of Grid-Forming Inverters in Unbalanced Distribution Grids," \textit{IEEE Transactions on Power Systems}, vol. 39, no. 2, pp. 4311-4322, March 2024.

\bibitem{b4}  E. Fayad, A. Bruyere, F. Colas and F. Gruson, "Single Phase Grid-Forming Control Implementation on Embedded Electric Vehicle Charger for V2X Applications," 2025 IEEE Kiel PowerTech, Kiel, Germany, 2025.

\bibitem{b5} S. Mossing, O. Amestegui, M. Jonas, f. Feng, L. Wang, Q. Shen, S. Zarrabian, Z. Liu, and P.  Zhang, "Generalized shipboard microgrid power flow incorporating hierarchical control," \textit{iEnergy}, vol. 4, no. 3, pp. 165-173, September 2025.

\end{thebibliography}
\end{document}